\documentclass[aps,prl,twocolumn,superscriptaddress]{revtex4}

\usepackage{graphicx}% Include figure files
\usepackage{bm}% bold math
\usepackage{color}

\begin{document}

%\title{London penetration depth and superfluid density in single crystals of Fe$_{1.03}$(Te$_{0.63}$Se$_{0.37}$) and Fe$_{1.06}$(Te$_{0.88}$S$_{0.14}$) superconductors}

\title{London penetration depth and superfluid density in single crystals of Fe(Te,Se) and Fe(Te,S) superconductors}

\author{H. Kim}
\affiliation{Ames Laboratory and Department of Physics \& Astronomy, Iowa State University, Ames, IA 50011}

\author{C. Martin}
\altaffiliation[Current Address: ]{Department of Physics, University of Florida, Gainesville, FL 32611}
\affiliation{Ames Laboratory and Department of Physics \& Astronomy, Iowa State University, Ames, IA 50011}

\author{R. T. Gordon}
\affiliation{Ames Laboratory and Department of Physics \& Astronomy, Iowa State University, Ames, IA 50011}

\author{M. A. Tanatar}
\affiliation{Ames Laboratory and Department of Physics \& Astronomy, Iowa State University, Ames, IA 50011}

\author{J. Hu}
\affiliation{Department of Physics and Engineering Physics, Tulane University, New Orleans, LA 70118}

\author{B. Qian}
\affiliation{Department of Physics and Engineering Physics, Tulane University, New Orleans, LA 70118}

\author{$\textmd{Z. Q. Mao}$}
\affiliation{Department of Physics and Engineering Physics, Tulane University, New Orleans, LA 70118}

\author{Rongwei Hu}
\altaffiliation[Current Address: ]{Ames Laboratory and Department of Physics \& Astronomy, Iowa State University, Ames, IA 50011}
\affiliation{Condensed Matter Physics and Materials Science Department, Brookhaven National Laboratory, Upton, NY 11973}

\author{C. Petrovic}
\affiliation{Condensed Matter Physics and Materials Science Department, Brookhaven National Laboratory, Upton, NY 11973}

\author{N. Salovich}
\affiliation{Loomis Laboratory of Physics, University of Illinois at Urbana-Champaign, 1110 West Green St. Urbana, IL 61801}

\author{R. Giannetta}
\affiliation{Loomis Laboratory of Physics, University of Illinois at Urbana-Champaign, 1110 West Green St. Urbana, IL 61801}

\author{R. Prozorov}
\email[Corresponding author: ]{prozorov@ameslab.gov}
\affiliation{Ames Laboratory and Department of Physics \& Astronomy, Iowa State University, Ames, IA 50011}

\date{12 January 2010}

\begin{abstract}
The in-plane London penetration depth, $\lambda(T)$, was measured in single crystals of the iron-chalcogenide superconductors Fe$_{1.03}$(Te$_{0.63}$Se$_{0.37}$) and Fe$_{1.06}$(Te$_{0.88}$S$_{0.14}$) by using a radio-frequency tunnel diode resonator. As is also the case for the iron-pnictides, these iron-chalcogenides exhibit a nearly quadratic temperature variation of $\lambda(T)$ at low temperatures. The absolute value of the penetration depth in the $T \to 0$ limit  was determined for Fe$_{1.03}$(Te$_{0.63}$Se$_{0.37})$ by using an Al coating technique, giving $\lambda(0)\approx560 \pm 20$ nm. The superfluid density $\rho_s(T)=\lambda^2(0)/\lambda^2(T)$ was fitted with a self-consistent two-gap $\gamma-$model. While two different gaps are needed to describe the full-range temperature variation of $\rho_s(T)$, a non-exponential behavior at low temperatures requires additional factors, such as scattering and/or significant gap anisotropy.
\end{abstract}

\pacs{74.25.Nf,74.20.Rp,74.20.Mn}
\maketitle

The majority of recently found iron-based superconductors are pnictides \cite{feas_YKamihara_HHosono}. The only exception so far is binary iron-chalcogenide FeSe \cite{fe_FHsu_MWu} that becomes superconducting with the excess Fe occupying interstitial sites of the (Te,Se) layers \cite{fe_WBao_ZMao}. In these materials, generally referred to as "11" compounds, Fe forms square planar sheets whereas Se ions form distorted tetrahedra surrounding the Fe ions, which is similar to the structure of the Fe-As pnictides. The electronic structure is also similar to pnictides. For "11" system it has been suggested both theoretically \cite{fe_ASubedi_MDu} and experimentally \cite{fe_TXia_QMZhang} that superconductivity could be magnetically mediated. Furthermore, the series of iron-chalcogenides from FeS through FeTe was theoretically explored within the spin-fluctuation picture, concluding that doped FeTe could exhibit the strongest superconductivity \cite{fe_ASubedi_MDu}. The systems over which the doping is most controlled are FeTe$_{1-x}$Se$_x$ \cite{fe_KYeh_MWu} and FeTe$_{1-x}$S$_x$ \cite{Mizuguchi2009}. So far the highest $T_c \approx15$ K is reported for the Fe(Te,Se) system \cite{fe_KYeh_MWu,fe_MFang_ZMao}.  The connection between superconductivity and magnetism in the "11" system has been demonstrated by the observation of the antiferromagnetic order in Fe$_{1+y}$Te \cite{fe_WBao_ZMao} and a spin resonance in Fe$_{1+y}$(Te$_{0.6}$Se$_{0.4}$) \cite{fe_YQiu_ZMao}. Fe(Te,S) is a superconductor with $T_c \approx 8.8$ K and its comprehensive characterization is described in Ref.\,\onlinecite{fe_RHu_CPetrovic}.

The "11" system exhibits many interesting phenomena. The transition temperature can be enhanced up to 37 K by applying modest pressures \cite{fe_SMargadonna_KPrassides}, which is comparable to the $T_c$ of iron-arsenide superconductors. The connection between $T_c$ and the pressure has been suggested to come from the enhancement of spin fluctuations \cite{fe_TImai_RCava} and from the modulation of electronic properties due to evolution of the inter-layer Se-Fe-Se separations \cite{fe_SMargadonna_KPrassides}. Several experimental works explore pairing mechanism of "11" compounds \cite{fe_HKotegawa_YTakano,fe_RKhasanov_NZhigadlo,fe_JDong_SLi}. The absence of a coherence peak in NMR measurements on polycrystalline FeSe suggests unconventional superconductivity \cite{fe_HKotegawa_YTakano}, while the power-law temperature dependence of the spin-relaxation rate, $1/T_1 \sim T^3$, could be reconciled with both a nodal gap or a fully-gapped $s_{\pm}$ state. Muon spin rotation study of the penetration depth in FeSe$_x$ was consistent with either anisotropic $s$-wave or a two-gap extended $s$-wave pairing \cite{fe_RKhasanov_NZhigadlo}. Thermal conductivity measurements concluded multigap nodeless superconductivity in polycrystalline FeSe$_x$ \cite{fe_JDong_SLi}.

%Precision measurements of the the magnetic penetration depth, $\lambda(T)$, are among the most useful tools to probe the symmetry of the superconducting gap.

In this work, we present an experimental study of the London penetration depth, $\lambda(T)$, in single crystals of Fe$_{1.03}$(Te$_{0.63}$Se$_{0.37}$) and Fe$_{1.06}$(Te$_{0.88}$S$_{0.14}$). We found that at low temperatures $\Delta \lambda(T)\propto T^n$ with $n\approx 2.1$ for Fe(Te,Se) and $n\approx 1.8$ for Fe(Te,S). The absolute value of $\lambda(0)\approx560$ nm was determined for Fe(Te,Se) by measuring the total $\lambda(T)$ of the sample coated with a thin Al film \cite{tdr_RProzorov_ABanks}. The in-plane superfluid density $\rho_s(T)=\lambda^2(0)/\lambda^2(T)$ was analyzed in the framework of a self-consistent two-gap $\gamma-$model \cite{feas_VKogan_RProzorov}.

% Figure 1.
\begin{figure}[tb]
\includegraphics[width=8 cm]{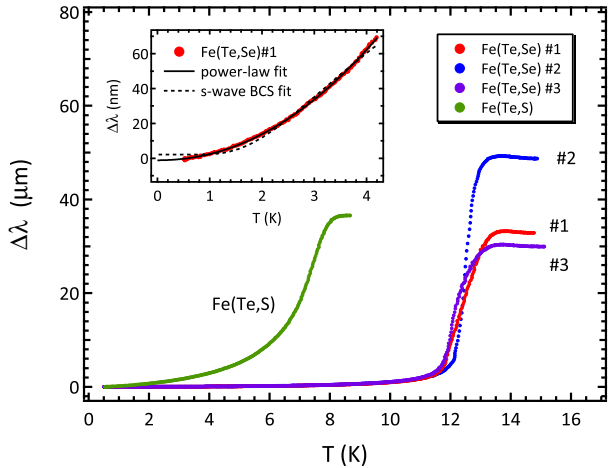}
\caption{(Color online) Main panel: Variation of the London penetration depth, $\Delta \lambda(T)$, for three Fe(Te,Se) samples and one Fe(Te,S) sample. Inset, $\Delta \lambda(T)$ for Fe(Te,Se) \#1 in the low temperature range up to $T_c/3$ shown along with the fitting curves assuming exponential or power-low behavior.}
\label{fig1}
\end{figure}

Single crystals of Fe(Te,Se) were synthesized using a flux method. Mixed powders of the Fe(Te$_{0.6}$Se$_{0.4}$) compositions were sealed in evacuated quartz tubes. The sealed ampoule was slowly heated up to 930 $^\circ$C and slowly cooled down to 400 $^\circ$C at a rate of 3 $^\circ$C/hr before the furnace was shut down. Single crystals with centimeter dimensions can easily be obtained with this method and are shown to be the pure $\alpha$-phase with the $P4/nmm$ space group by x-ray diffraction \cite{fe_MFang_ZMao}. The actual concentrations were analyzed using an energy dispersive x-ray spectrometer (EDXS). The measured composition for the samples used in this work is Fe$_{1.03}$(Te$_{0.63}$Se$_{0.37}$). Single crystals of Fe(Te,S) were grown from TeS self flux using a high temperature flux method. Elemental Fe, Te and S were sealed in quartz tubes under a partial argon atmosphere. The sealed ampoule was heated to a soaking temperature of 430-450 $^\circ$C for 24h, followed by a rapid heating to the growth temperature at 850 $^\circ$C and then slowly cooled to 820 $^\circ$C. The excess flux was removed from the crystals by decanting. The end composition of the sample used to measure the penetration depth was Fe$_{1.06}$(Te$_{0.88}$S$_{0.14}$)\cite{fe_RHu_CPetrovic}.

%Actual concentration for s1 = (Fe=1.001+-0.00241,Te=0.632+-0.00382,Se=0.366+-0.00387), s2=(Fe=1.002+-0.00241,Te=0.638+-0.00395,Se=0.360+-0.00432)

The in-plane London penetration depth, $\lambda (T)$, was measured by using a self-oscillating tunnel-diode resonator (TDR) \cite{tdr_CDegrift,tdr_RProzorov_FAraujo,tdr_RProzorov_ABanks,tdr_RProzorov_WGiannetta}. The sample was mounted on a sapphire rod which was inserted into a tank-circuit inductor. The weak ac magnetic field $H_{ac} \sim 20$ mOe produced by the coil is much smaller than the lower critical field $H_{c1} < 100$ Oe, so the sample was in the Meissner state and its magnetic response was determined by the London penetration depth. To probe the $ab-$plane supercurrent response, the sample was placed with its crystallographic $c-$axis along $H_{ac}$. The shift of the resonance frequency, $\Delta f \equiv f(T)-f_0$, is measured to obtain the total magnetic susceptibility $\chi (T)$ via $\Delta f = -G4\pi\chi (T)$. Here $f_0 \sim 14$ MHz is the resonance frequency of an empty resonator, $G=f_0V_s/2V_c(1-N)$ is the calibration factor that depends on the demagnetization factor $N$, sample volume $V_s$ and coil volume $V_c$. The calibration factor is determined for each sample by measuring the full frequency change resulting from physically pulling the sample out of the coil at the lowest temperature. In the Meissner state the magnetic susceptibility, $4 \pi \chi$, can be written in terms of $\lambda$ and the effective sample dimension $R$ as: $4\pi\chi=(\lambda /R) \tanh (R/\lambda)-1$, from which $\lambda$ can be obtained \cite{tdr_RProzorov_FAraujo}.

% Figure 2.
\begin{figure}[tb]
\includegraphics[width=8 cm]{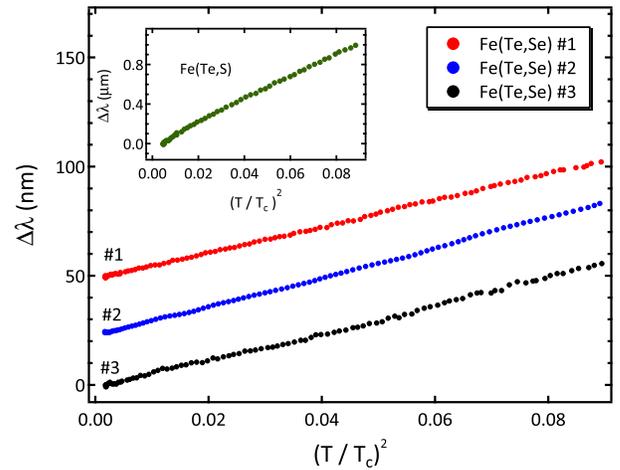}
\caption{(Color online) $\Delta \lambda$ plotted vs. $(T/T_c)^2$ for three Fe(Te,Se) crystals in the main panel and Fe(Te,S) crystal in the inset in the temperature range up to $T_c/3$. The curves for Fe(Te,Se)\#1 and \#2 are shifted vertically for clarity by 50 nm and 25 nm, respectively.}
\label{fig2}
\end{figure}

The main panel in Fig.\,\ref{fig1} shows the full-temperature range penetration depth for Fe(Te,Se) and Fe(Te,S) superconductors. While the Fe(Te,Se) samples show a fairly sharp superconducting transition with $\Delta T_c < 1.5$ K reflecting the high quality of the single crystals, the Fe(Te,S) crystal shows a broader transition. The "maximum slope", $T^\textmd{\scriptsize slope}_c$, determined by taking the maximum of the derivative $d\Delta\lambda (T)/dT$ gives $T^\textmd{\scriptsize slope}_c\approx 12.0$ K for Fe(Te,Se) and $T^\textmd{\scriptsize slope}_c\approx 7.5$ K for Fe(Te,S). As for the onset values,  $T^\textmd{\scriptsize onset}_c \approx 13$ K for Fe(Te,Se) and $T^\textmd{\scriptsize onset}_c \approx 8$ K for Fe(Te,S). The low-temperature variation of $\lambda(T)$ is examined in the inset in Fig.~\ref{fig1}. The dashed line represents the best fit to a standard $s$-wave BCS function, $\Delta \lambda(T)=\lambda(0)\sqrt{\pi \Delta_0/2T}\exp(-\Delta_0/T)$, with $\lambda(0)$ and $\Delta_0$ being free fitting parameters. The experimental data do not show any indication of saturation down to $0.04T_c$ and the fit is not adequate. Also obtained from the fit is $\Delta_0 = 0.5 T_c$, which is impossible in a single-gap scenario, hence ruling out conventional $s$-wave BCS superconductivity. We will come back to a multi-gap $s$-wave fitting later in the discussion of the superfluid density. On the other hand, fitting with the power-law, $\Delta \lambda (T)\propto A T^n$, $n=2.1\pm0.01$, produces excellent agreement with the data.

% Figure 3
\begin{figure}[tb]
\includegraphics[width=8 cm]{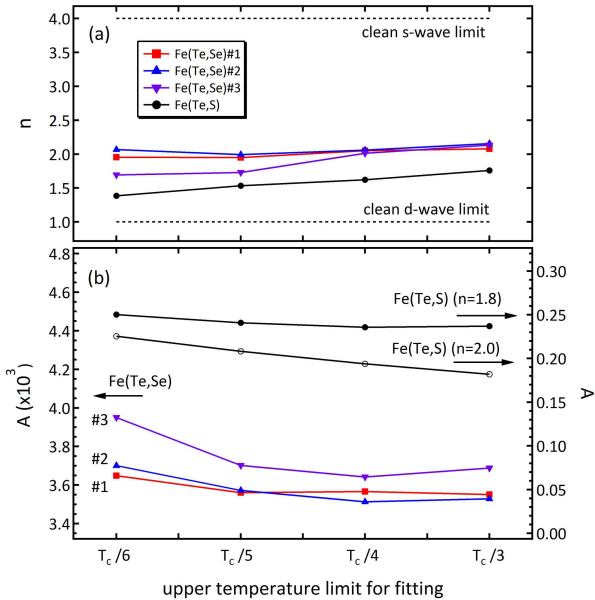}
\caption{(Color online) Exponent $n$ and pre-factor $A$ obtained by fitting to $\Delta \lambda (T)\propto A T^n$ for various upper temperature limits shown on the x-axis. The exponents in the upper panel were obtained with $n$ and $A$ both being free parameters. In the lower panel, $A$ was acquired with a fixed $n$.}
\label{fig3}
\end{figure}

In order to examine how close the overall power-law variation is to quadratic, we plot $\Delta\lambda$ versus $(T/T_c)^2$ for Fe(Te,Se) in the main panel of Fig.\,\ref{fig2} and for Fe(Te,S) in the inset. All samples follow the $\Delta \lambda(T)\propto T^2$ behavior rather well. To probe how robust the power $n$ is, we performed a data fit over a floating temperature range, from $T$=0 to $T_\textmd{\scriptsize up}$, using a functional form of $\Delta\lambda(T)= a_0+ A T^n$. The difference between the $a_0$ term determined from an extrapolation from the $T^2$ plot in Fig.~\ref{fig2} and the power-law fit turned out to be negligible, 1.5$\pm$0.5 nm, and had no significant effect on the fit. The dependence of the other fitting parameters, $n$ and $A$, on $T_\textmd{\scriptsize up}$ (selected in the range from $T_c/6$ to $T_c/3$) is summarized in Fig.\,\ref{fig3}. The upper panel of Fig.\,\ref{fig3} shows the exponent $n$, which (i) does not depend much on the selection of the upper limit of the fitting range, (2)  is somewhat smaller in Fe(Te,S), $n \approx 1.8$, than in Fe(Te,Se) $n \approx 2$. The  pre-factor $A$ obtained from the fit does not depend much on the fitting range either.

\begin{figure}[tb]
\includegraphics[width=8cm]{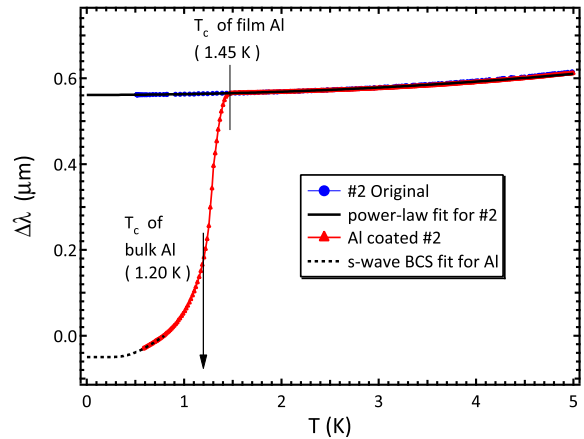}
\caption{(Color online) Effective penetration depth in single crystal Fe(Te,Se) before (blue circles) and after (red triangles) coating with an Al layer. The curve is shifted up according to Eq.\,(\ref{L0}) and the data are extrapolated to $T=0$ using a $T^2$ fit resulting in $\lambda(0)\approx 560 \pm 20$ nm.}
\label{fig4}
\end{figure}

% the absolute value of lambda using Al coating
To calculate the superfluid density, we need to know the absolute value of the penetration depth, $\lambda(0)$.  We used the technique described in Ref.\,\onlinecite{tdr_RProzorov_ABanks}. A thin aluminum layer was deposited using magnetron sputtering conducted in an argon atmosphere. The Al layer thickness, $t =100 \pm 10$ nm, was determined by using an Inficon XTC 2 with a 6 MHz gold quartz crystal and later directly measured by using scanning electron microscopy on the edge of a broken sample. By measuring the frequency shift from $T \ll T^\textmd{\scriptsize Al}_c$ to $T>T^\textmd{\scriptsize Al}_c$ and converting it into the effective penetration depth of the coated sample, $\lambda_\textmd{\scriptsize eff}$, one can extract the full penetration depth of the material under study from

\begin{equation}
\lambda_\textmd{\scriptsize eff}=\lambda_\textmd{\scriptsize Al}\frac{\lambda+\lambda_\textmd{\scriptsize Al}\tanh{(t/\lambda_\textmd{\scriptsize Al})}}{\lambda_\textmd{\scriptsize Al}+\lambda\tanh{(t/\lambda_\textmd{\scriptsize Al})}}
\label{L0}
\end{equation}

\noindent where $\lambda$ is the unknown penetration depth to be determined. Figure \ref{fig4} shows
the measured $\lambda_\textmd{\scriptsize eff}(T)$ that is compared to the data without Al coating. The negative offset of 0.05 $\mu$m accounts for the thickness of the Al layer and $\lambda_{Al}(T \ll T^\textmd{\scriptsize Al}_c)$. According to Eq.\,(\ref{L0}), data plotted this way give the actual $\lambda(T)$ and its extrapolation to $T=0$ gives an estimate of $\lambda(0)\approx 560 \pm 20$ nm for the penetration depth of Fe(Te,Se). More details on the method can be found in Ref.\,\onlinecite{tdr_RProzorov_ABanks}.

% Figure 4 absolute value measurement

\begin{figure}[tb]
\includegraphics[width=8cm]{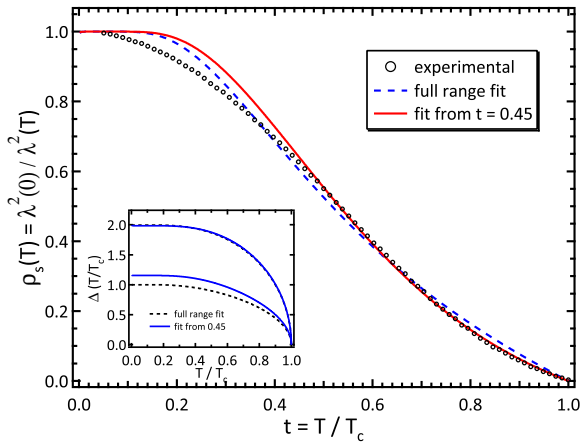}
\caption{(Color online) Superfluid density $\rho_s(T/T_c)$ for Fe(Te,Se)\#2 calculated with experimental $\Delta \lambda(T)$ and $\lambda (0)=560$ nm. The solid (red) line is a fit the two-gap $\gamma-$model from $0.45 T_c$ to $T_c$. The dashed (blue) line is the fit over the full temperature range. Inset: temperature dependent superconducting gaps calculated self-consistently during the fitting.}
\label{fig5}
\end{figure}

The superfluid density, $\rho_s(T)=\lambda^2(0)/\lambda^2(T)$, shown in Fig.\,\ref{fig5}, exhibits a noticeable positive curvature at elevated temperatures, similar to MgB$_2$\cite{mgb2_JFletcher_JKarpinski}. This suggests a multi-gap superconductivity, which we analyze in the framework of the self-consistent $\gamma-$model \cite{feas_VKogan_RProzorov}. Although this $s-$wave model cannot explain the power-law behavior at the lowest temperatures, it should provide a reasonable description at elevated temperatures, $\sim T_c/3 <T <T_c$,  for which effects of gap anisotropy are smeared. Fitting in the temperature range from $0.45 T_c$ to $T_c$, shown by a solid (red) line in Fig.\,\ref{fig5}, produces a good agreement with the data. To limit the number of the fitting parameters, the partial densities of states were chosen to be equal in the two bands, $n_1=0.5$, and the first intra-band coupling parameter, $\lambda_1=0.5$, was chosen to produce a correct $T_c \approx 12$ K assuming a Debye temperature of 400 K. The variation of $\lambda_1$ does not affect the fitting quality or relative ratios of the fitting parameters. The parameters obtained in the fit are: $\lambda_2=0.347$, $\lambda_{12}=0.096$ and $\gamma =  0$. This result means that $\rho_s (T)$ at temperatures of the order of $T_c$ is fully described by only one component, determined by the band with a smaller gap. The existence of the larger gap and small interband coupling, $\lambda_{12}$, are needed, however, to maintain a high $T_c$. The fit over the entire temperature range reveals a clear deviation from this clean exponential model at low temperatures. The new fitting parameters of $\lambda_2=0.281$, $\lambda_{12}=0.117$ and $\gamma =  0.157$ are close to the previous set, albeit with small, but finite $\gamma$ indicating 16 \% contribution of the larger gap to the total superfluid density. The temperature dependent gaps obtained self-consistently in the fitting are shown in the inset to Fig.\,\ref{fig5}. While the fitted positive curvature and reasonable coupling parameters indicate a multi-gap nature of superconductivity in "11" iron-chalcogenide superconductors, the failure at low temperatures and apparently non-exponential behavior requires extension to the anisotropic gap and inclusion effects of (possibly strong) pairbreaking \cite{Gordon2010}.

% Figure 5

In conclusion, a robust power-law behavior of the low-temperature $\lambda(T)$ is found in single crystals of Fe$_{1.03}$(Te$_{0.63}$Se$_{0.37}$) and Fe$_{1.06}$(Te$_{0.88}$S$_{0.14}$) with $\Delta\lambda(T) \propto T^n$ with $n\approx 2.1$ and $\approx 1.8$, respectively. For Fe(Te,Se), the absolute value, $\lambda(0)\approx 560 \pm 20 $ nm, was determined by the coating technique. The analysis of the superfluid density shows a clear signature of multi-gap superconductivity and a failure of the clean limit s-wave (including $s_{\pm}$) pairing.

%Acknowledgement
We thank V. G. Kogan for helpful discussions. Work at the Ames Laboratory was supported by the Department of Energy-Basic Energy Sciences under Contract No. DE-AC02-07CH11358. R. P. acknowledges support from the Alfred P. Sloan Foundation. The works at Tulane and UIUC are supported by the NSF under grant DMR-0645305 and DMR-0503882, respectively, and part of this research was carried out at Brookhaven National Laboratory, which is operated for the U.S. Department of Energy by Brookhaven Science Associates (DE-AC02-98CH10886)

\end{document}